\documentclass[aip, jcp, reprint]{revtex4-1}
\usepackage{graphicx}
\usepackage{amsmath}
\usepackage{amssymb}
\usepackage{color}
\usepackage{gensymb}

\newcommand{\nto}[0]{$\mathrm{N_2O}$}
\newcommand{\cst}[0]{$\mathrm{CS_2}$}
\newcommand{\ocs}[0]{$\mathrm{OCS}$}

\newcommand{\Trev}{{T_\text{rev}}}
\newcommand{\Jd}{J_\text{D}}
\newcommand{\Jc}{J_\text{C}}

\begin{document}

\title{Probing molecular potentials with an optical centrifuge}

\author{A. A. Milner}
\author{A. Korobenko}
\affiliation{Department of  Physics \& Astronomy, The University of British Columbia, Vancouver, Canada}
\author{J. W. Hepburn}
\affiliation{Department of Chemistry, The University of British Columbia, Vancouver, Canada}
\author{V. Milner}
\email{vmilner@phas.ubc.ca}
\affiliation{Department of  Physics \& Astronomy, The University of British Columbia, Vancouver, Canada}

\date{\today}

\begin{abstract}
        We use an optical centrifuge to excite coherent rotational wave packets in \nto{}, \cst{} and \ocs{} molecules with rotational quantum numbers reaching up to $J\approx 465$, 690 and 1186, respectively. Time-resolved rotational spectroscopy at such ultra-high levels of rotational excitation can be used as a sensitive tool to probe the molecular potential energy surface at internuclear distances far from their equilibrium values. Significant bond stretching in the centrifuged molecules results in the growing period of the rotational revivals, which are experimentally detected using coherent Raman scattering. We measure the revival period as a function of the centrifuge-induced rotational frequency and compare it with the numerical calculations based on the known Morse-cosine potentials.
    \end{abstract}

\maketitle

Potential energy surfaces (PES) are central to almost every aspect of molecular dynamics as they govern both the motion of atoms inside a single molecule \cite{Green1992} as well as collisions and chemical reactions between several molecular reagents \cite{Polanyi1995}. Mapping out PES of polyatomic molecules, possessing many degrees of freedom, is a difficult task which becomes especially challenging far from the equilibrium, where multiple vibrational and rotational modes become strongly coupled. A common approach consists of expressing the ro-vibrational energy of a molecule as a function of internal molecular coordinates, calculate the frequencies of vibrational transitions and adjust the expansion coefficients so as to match the calculated frequencies with experimental observations. The higher the energies of the observed ro-vibrational transition, the wider the region of the potential energy surface which can be probed and accurately mapped out, the richer the molecular dynamics available for quantitative studies.

A great deal of work has been done on triatomic molecules, for which precise spectroscopic information about highly excited vibrational transitions is readily available. For the molecules studied here -- \nto{}, \ocs{} and \cst{}, vibrational energies up to approximately $15\times 10^{3}$~cm$^{-1}$, $13\times 10^{3}$~cm$^{-1}$ and $14\times 10^{3}$~cm$^{-1}$, respectively, have been reported. In a series of works, Z\'{u}\~{n}iga \textit{et al.} \cite{Zuniga1999, Zuniga2000, Zuniga2002} developed optimal generalized internal coordinates to fit the ro-vibrational spectrum of these molecules by approximating their potential energy surface with a fourth-order Morse-cosine expansion \cite{Jensen1988}:
\begin{eqnarray}\label{Eq-MorseCosine}
V(R_1, R_2, \theta ) &=& M_{2} y_{2} + \sum_{ij} M_{ij}y_{i}y_{j}+ \sum_{ijk} M_{ijk}y_{i}y_{j} y_{k} \nonumber \\
&+& \sum_{ijkl} M_{ijkl} y_{i}y_{j}y_{k}y_{l},
\end{eqnarray}
where $R_1$ and $R_2$ are the lengths of the two molecular bonds, $\theta $ is the angle between them, $M$ are the expansion coefficients and
\begin{eqnarray}\label{Eq-MorseCosineYs}
y_{1} &=& 1 - e^{-a_{1} (R_{1}-R_{1,0})}, \nonumber \\
y_{2} &=& \cos \theta - \cos \theta_{0}, \\
y_{3} &=& 1 - e^{-a_{3} (R_{2}-R_{2,0})}, \nonumber
\end{eqnarray}
with the subscript `0' denoting the equilibrium value for the corresponding coordinate in the ground ro-vibrational state. Constants $a_{1}$ and $a_{3}$ determine the stiffness of the respective bonds.

\begin{figure}[t]
    \includegraphics[width=.8\columnwidth]{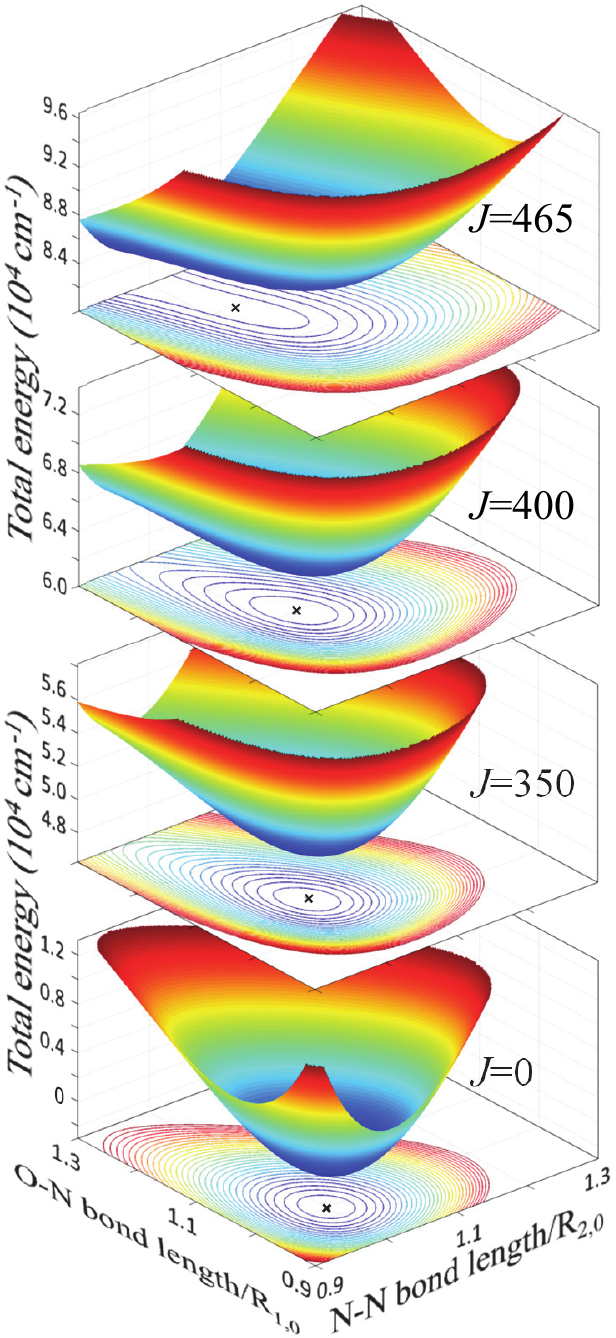}
    \caption{(color online) Calculated potential energy surfaces of \nto{} at four values of the molecular angular momentum $J=0, 350, 400$ and 465. The latter represents the largest value of $J$, for which the PES still has a minimum. The lengths of O--N and N--N bonds along the two horizontal axes are normalized to the corresponding equilibrium distances $R_{1,0}$ and $R_{2,0}$. Black crosses mark the position of the energy minimum $(R_{1,J},R_{2,J})$. Note the increasing energy scale.}
    \label{Fig-PES}
\end{figure}

In this work, we use high-energy rotational, rather than vibrational, excitation to study the molecular potentials at the limit when the molecular bonds are stretched far beyond their ground-state values. We employ the technique of an optical centrifuge \cite{Karczmarek1999, Villeneuve2000} to spin the molecules of interest to ultra-high states of angular momentum $J$. The total energy of a rotating molecule can be expressed as
\begin{equation}\label{Eq-Energy}
E_{J}(R_{1}, R_{2}, \theta )= V(R_1, R_2, \theta )+\frac{\hbar^2 J (J+1)}{2\mathbf{I}(R_1, R_2, \theta )},
\end{equation}
where $\mathbf{I}$ is the molecular moment of inertia and $\hbar$ is the reduced Planck's constant. As the angular momentum increases, the second term in the above expression pushes up the equilibrium bond lengths $R_{1,J}$ and $R_{2,J}$, corresponding to the minimum of the total energy, to higher and higher values, eventually resulting in dissociation. The effect is illustrated in Figure~\ref{Fig-PES}, where we plot the potential energy surfaces of \nto{} at different levels of rotational excitation, calculated with Eq.\ref{Eq-Energy} and the Morse-cosine PES coefficients from Ref.~\citenum{Zuniga1999}. While not much distortion is happening below $J\approx 300$, the shape of the PES changes rapidly at higher values of the molecular angular momentum, until the potential minimum disappears at the dissociation limit $\Jd=465$. Notably, before the molecule dissociates, its total energy increases to about $10^{5}$~cm$^{-1}$, almost an order of magnitude above the level typically accessible through vibrational excitation \cite{Campargue1995}.

Rotational dissociation by an optical centrifuge has been demonstrated experimentally with a diatomic chlorine \cite{Villeneuve2000}, and theoretically studied for both the diatomic (Cl$_{2}$, Ref.\citenum{Karczmarek1999, Spanner2001}) and triatomic (HCN, Ref. \citenum{Hasbani2002}) molecules. In both cases, it was predicted that the majority of molecules undergoing forced accelerated rotation will be ejected from the centrifuge prior to reaching the dissociation energy due to the Coriolis force. The latter causes the molecular axis to turn away from the laser field polarization and eventually fall behind the rotating trap. Here, we analyze the molecular dynamics in the vicinity of this \textit{Coriolis wall} both numerically and experimentally, in several centrifuged triatomic molecules.

\begin{figure}[t]
    \includegraphics[width=.7\columnwidth]{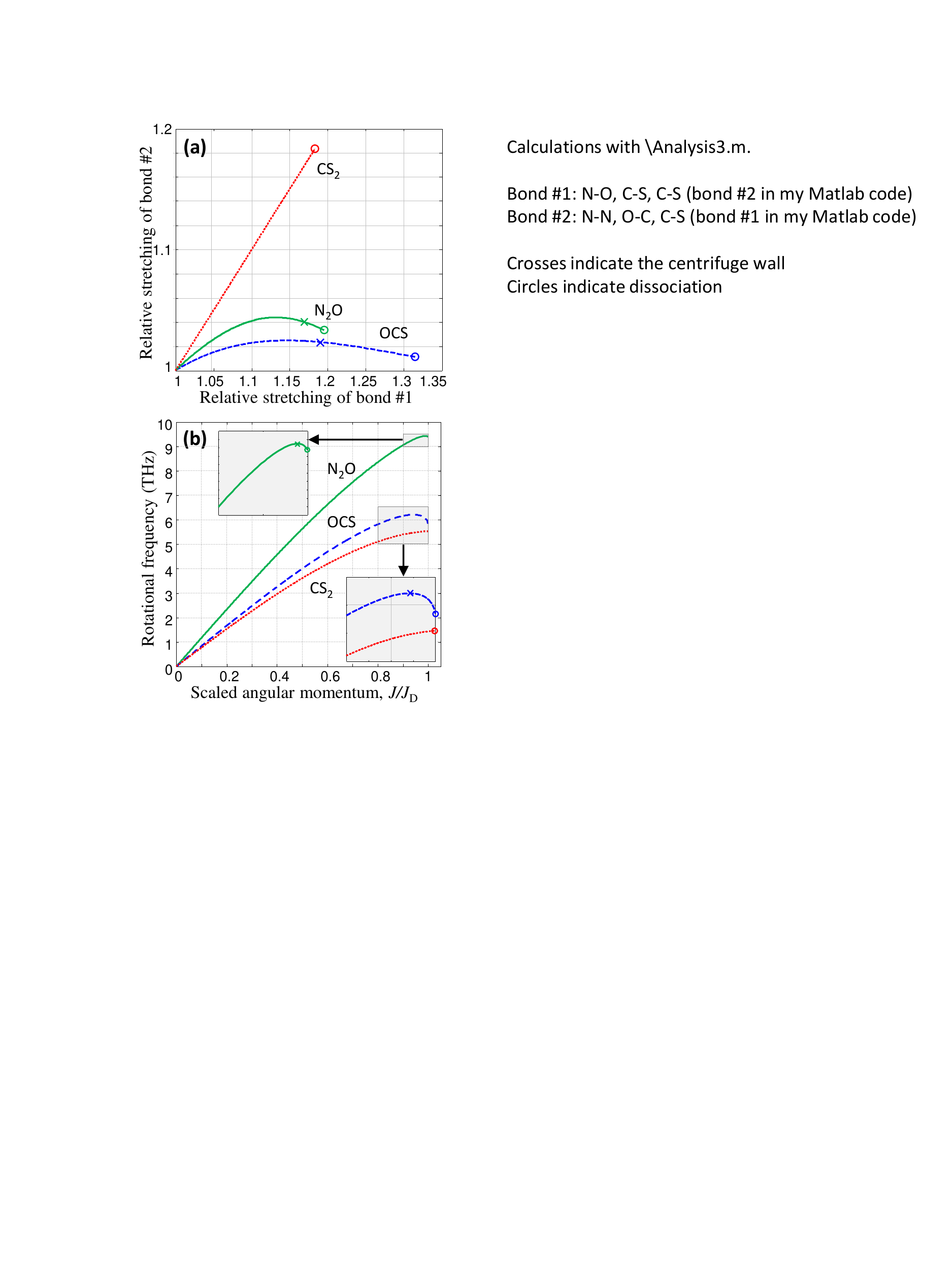}
    \caption{(color online) (\textbf{a}) Relative stretching of bond \#1 (O--N, S--C, S--C) and bond \#2 (N--N, O--C, S--C) of \nto{}, \ocs{} and \cst{}, respectively, plotted in units of the corresponding equilibrium distances in the ground ro-vibrational state. The angular momentum is increasing along each curve from $J=0$ (lower left corner) to the dissociation limit $\Jd$, labeled with circles. (\textbf{b}) Classical rotational frequency of \nto{}, \ocs{} and \cst{} as a function of the molecular angular momentum $J$, normalized by its value at dissociation ($\Jd$). In both panels and insets, crosses mark the location of the Coriolis wall, which limits the adiabatic excitation by the centrifuge (see text for details).}
    \label{Fig-Stretch}
\end{figure}

Figure~\ref{Fig-Stretch}(\textbf{a}) shows the calculated stretching of both molecular bonds in \nto{}, \ocs{} and \cst{} with the rotational quantum number increasing from $J=0$ to the calculated dissociation limit $\Jd{}=465, 690$ and 1186, respectively. The corresponding dissociation energies are $83\times10^{3}$ cm$^{-1}$, $86\times10^{3}$ cm$^{-1}$ and $132\times10^{3}$ cm$^{-1}$. As anticipated for a linear molecule, the equilibrium angle between the bonds remains zero regardless of $J$. The main goal of this work is to demonstrate how such bond stretching under ultrafast centrifuge spinning can be used for probing high-energy regions of the potential energy surfaces. Although neither internuclear distances nor angular momenta are measured directly in our experiments, we accomplish the task by investigating the periodic dynamics of the centrifuged molecules as a function of the rotational frequency, controlled by the centrifuge. Our analysis is carried out as follows.

Given the rotational energy spectrum $E(J)$, the classical rotational frequency at any $J$ can be calculated as
\begin{equation}\label{Eq-Freq}
\nu _{J}= \frac{dE(J)/dJ}{2\pi\hbar}=\frac{E(J+1)-E(J)}{2\pi\hbar}.
\end{equation}
The results of these calculations in the simplest approximation which neglects the zero-point vibrational energy, i.e. $E(J)\approx E_J(R_{1,J}, R_{2,J}, 0)$, are shown in Fig.\ref{Fig-Stretch}(\textbf{b}). The growth of $\nu _{J}$ with the molecular angular momentum is sub-linear, indicating strong centrifugal distortion. Similarly to the previously discussed \cite{Karczmarek1999, Hasbani2002} behavior of Cl$_{2}$ and HCN, the rotational frequencies of \nto{} and \ocs{} fall off above the critical values of $\Jc=460$ (green solid) and $\Jc=650$ (dashed blue), respectively. The corresponding frequency maxima of $\nu _{C}=9.4$~THz and 6.2~THz, marked with crosses in Fig.\ref{Fig-Stretch}(\textbf{b}), represent the Coriolis wall on the way of accelerated molecular rotation. Beyond this wall, the distance between consecutive rotational levels starts decreasing and the molecules can no longer adiabatically follow the centrifuge field.
\begin{figure}[t]
    \includegraphics[width=.8\columnwidth]{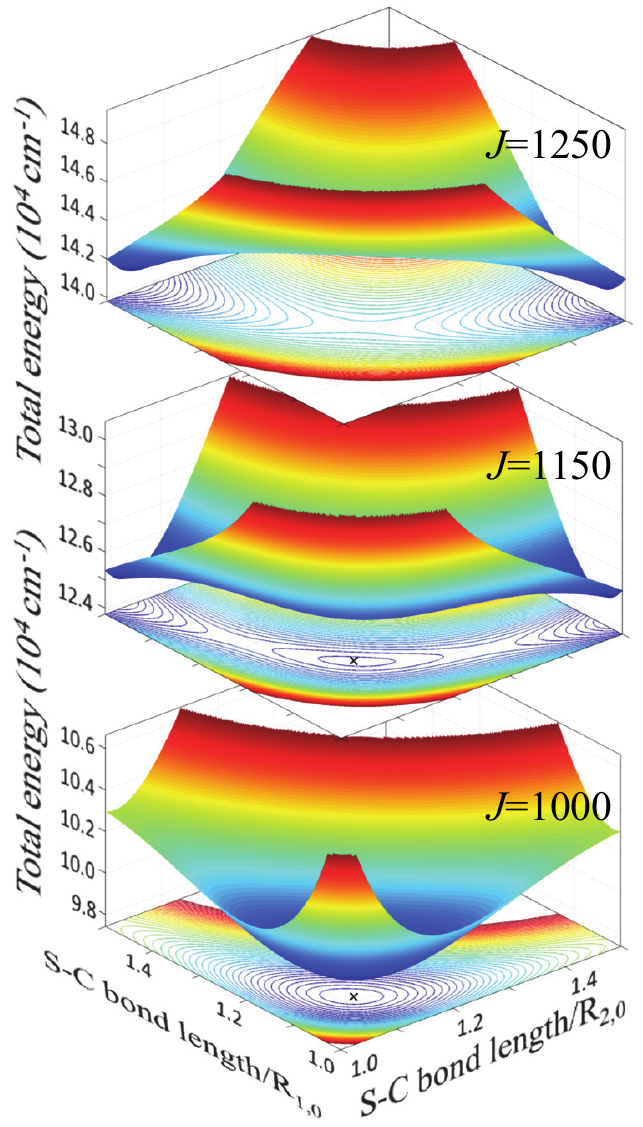}
    \caption{(color online) Calculated potential energy surfaces of \cst{} at  $J=1000, 1150$ and 1250, around the dissociation point of $\Jd=1186$. The lengths of both S--C bonds along the two horizontal axes are normalized to the equilibrium distance $R_{1,0}=R_{2,0}=1.55$ {\AA}. Black crosses in the two bottom panels mark the position of the energy minimum. Note the increasing energy scale.}
    \label{Fig-CS2}
\end{figure}

Interestingly, the rotational frequency of \cst{} increases monotonically (dashed red line), indicating that the molecule can climb the rotational ladder all the way to the dissociation frequency $\nu _{D}=5.5$~THz without escaping the centrifuge. This conclusion, supported by our experimental observations (discussed below), can be interpreted as follows. In asymmetric linear triatomic rotors, the centrifugal force stretches predominantly one of the two bonds (O--N, S--C and N--C in \nto{}, \ocs{} and HCN, respectively), whereas the other one extends to a much lesser degree and even starts shrinking with the increasing rotational frequency [see Fig.\ref{Fig-Stretch}(\textbf{a})]. The quickly stretching bond is responsible for the strong Coriolis force, which makes the molecule lag behind, and eventually fall off, the centrifuge without dissociating.

The symmetry of \cst{}, on the other hand, effectively stiffens the molecule with respect to the centrifugal pull. The lengths of both S--C bonds increase slower than in the asymmetric case, which results in the weaker Coriolis force and better stability in the centrifuge. The difference between the symmetric and asymmetric rotors can be better appreciated by comparing the respective potential energy surfaces near dissociation. While the Gaussian curvature (the product of two principal curvatures) of the potential well of \nto{} remains positive with increasing angular momentum (Fig.\ref{Fig-PES}), its sign changes abruptly when the PES of \cst{} develops saddle geometry at $J=1186$ (Fig.\ref{Fig-CS2}). The subsequent sudden elongation of one of the S--C bonds, even if accompanied by strong Coriolis force, leads to an inevitable dissociation of the molecule. Although the above classical picture does not account for the quantum tunneling out of the well, the latter will only expedite the dissociation process rather than prevent it.

Since the critical frequency values ($\nu _{C}$ and $\nu _{D}$) are within the reach of our optical centrifuge ($\leq 10$~THz), we were able to study the described behavior experimentally. After the molecules are spun up and released from the centrifuge, we monitor their dynamics by means of coherent Raman scattering \cite{Korobenko2014a}. An excited rotational wave packet undergoes periodic revivals, separated in time by \cite{Leichtle1996, Seideman1999}
\begin{equation}\label{Eq-Trev}
\Trev=\frac{ 2\pi \hbar} {d^2E(J)/dJ^2}= \frac{2\pi \hbar } {E(J-1)-2E(J)+E(J+1)}.
\end{equation}
Fractional revivals may also occur (as is the case here), depending on the particular structure of the wave packet \cite{Averbukh1989}. By comparing the experimentally found dependence of the revival time on the rotational frequency, $\Trev(\nu_{J})$, with the one calculated using the theoretical energy spectrum $E(J)$, the validity of the latter can be readily assessed.
\begin{figure}[t]
    \includegraphics[width=.7\columnwidth]{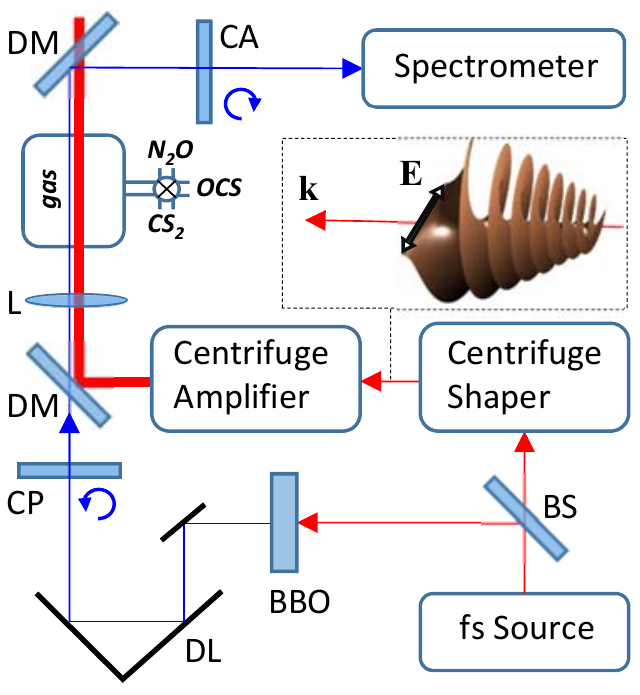}
    \caption{(color online) Experimental set up. BS: beam splitter, DM: dichroic mirror, CP/CA: circular polarizer/analyzer, DL: delay line, L: lens. The gas cell is filled with either \nto{}, \ocs{} or \cst{} under room temperature and pressure of 30~kPa, 10~kPa and 5~kPa, respectively. An optical centrifuge field is illustrated above the centrifuge shaper with \textbf{k} being the propagation direction and \textbf{E} the vector of linear polarization undergoing an accelerated rotation.}
    \label{Fig-Setup}
\end{figure}

\begin{figure}[t]
    \includegraphics[width=.7\columnwidth]{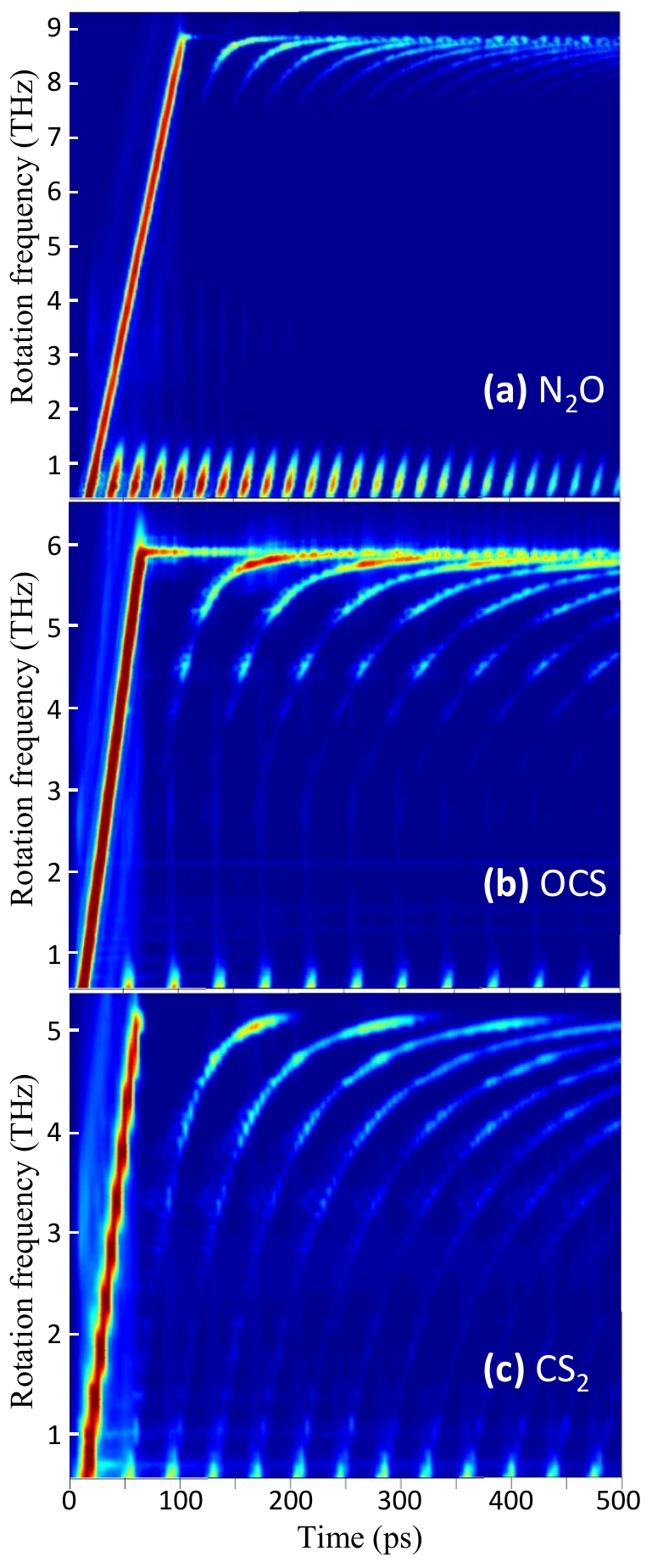}
    \caption{(color online) Time-resolved rotational Raman spectra of centrifuged (\textbf{a}) \nto{}, (\textbf{b}) \ocs{} and (\textbf{c}) \cst{} molecules as a function of the time delay between the centrifuge and probe pulses. Bright tilted lines correspond to the Raman signal from the coherently rotating molecules. Dim vertical ``shadow'' traces originating from the regions of strong Raman signal [e.g. between the frequencies of 1 and 3.5~THz, or above 6 THz in plot (\textbf{b})] are spectrometer's artifacts. See text for more details.}
    \label{Fig-Raman}
\end{figure}

The experimental setup, shown in Fig.~\ref{Fig-Setup}, is similar to that used in our previous work \cite{Korobenko2014a}. A beam of femtosecond pulses from a regenerative Ti:Sapphire amplifier (800~nm, 1~KHz repetition rate, 30~nm full width at half maximum) is split in two parts. One part is sent to a ``centrifuge shaper'', implemented according to the original recipe of Karczmarek \textit{et al.} \cite{Karczmarek1999}, which converts the input laser field into the field of an optical centrifuge, schematically illustrated in the inset. The shaper is followed by a home built Ti:Sapphire multi-pass amplifier (10~Hz repetition rate), which boosts the energy of centrifuge pulses up to 30~mJ/pulse. The pulses are about 100 ps long and their linear polarization undergoes an accelerated rotation, reaching the angular frequency of 10 THz by the end of the pulse.  The second (probe) beam is frequency doubled in a nonlinear BaB$_2$O$_4$ (BBO) crystal, time delayed by means of a controllable translation stage and combined with the centrifuge beam on a dichroic mirror. Both beams are focused into a cell, filled with either \nto{}, \ocs{} or \cst{} gas, with a $f=1$~m lens.
\begin{figure*}[]
    \includegraphics[width=1.8\columnwidth]{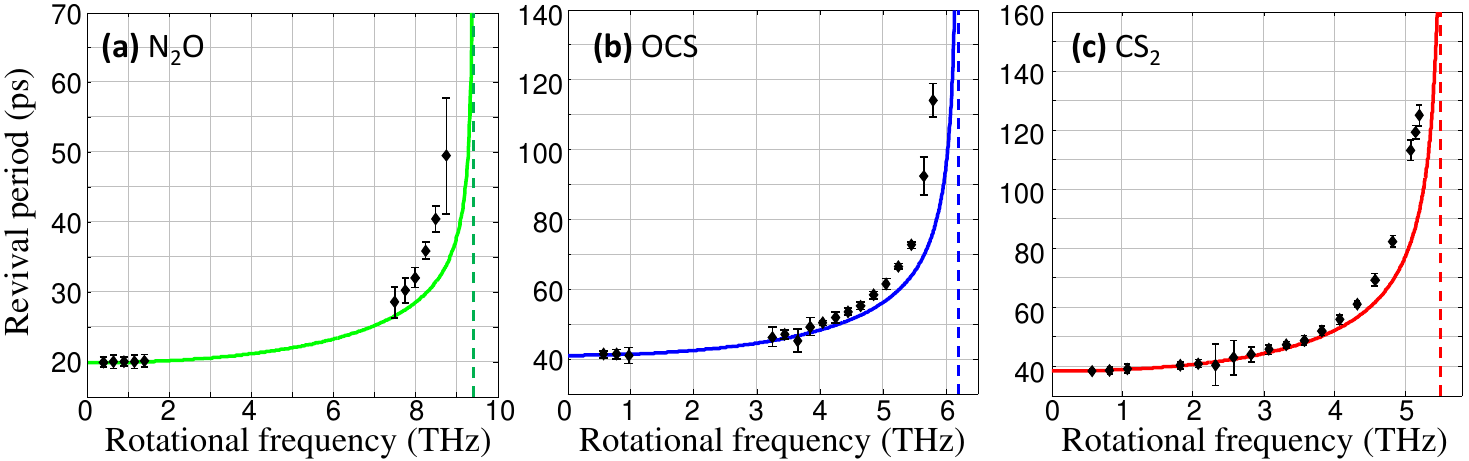}
    \caption{(color online) Revival periods extracted from the experimental data shown in Fig.\ref{Fig-Raman} (black diamonds with error bars), and theoretically calculated (solid lines) using Eq.\ref{Eq-Trev} with the total energy $E_{J}$ found from Morse-cosine potential energy surface, as explained in text. Dashed vertical lines mark the expected location of the Coriolis wall for \nto{} and \ocs{}, and the calculated dissociation frequency of \cst{}.}
    \label{Fig-Analysis}
\end{figure*}

We detect the dynamics of the centrifuged molecules using time-resolved coherent Raman spectroscopy. The centrifuge-induced coherence between the rotational states $|J,M=J\rangle$ and $|J+2,M=J+2\rangle$ (where $M$ is the projection of $\mathbf{J}$ on the propagation direction of the centrifuge field) results in the Raman frequency shift of the probe field. From the selection rules $\Delta M=2$ and the conservation of angular momentum, it follows that the Raman sideband of a circularly polarized probe is also circularly polarized, but with an opposite handedness. Due to this change of polarization, the strong background of the input probe light can be efficiently suppressed by means of a circular analyzer, orthogonal to the input circular polarizer (CA and CP, respectively, in Fig.\ref{Fig-Setup}). Raman spectra of the probe pulses scattered from the rotating molecules have been measured as a function of the probe delay relative to the centrifuge with a $f/4.8$ spectrometer equipped with a 2400 grooves/mm grating.

The experimentally measured Raman spectrograms are shown in Fig.\ref{Fig-Raman}. The bright tilted line on the left side of each plot reflects the accelerated rotation of molecules inside the centrifuge (in all three cases, the angular acceleration is exactly the same, whereas the tilt appears different due to the difference in vertical scales). While spinning up, the molecules are ``leaking'' from the centrifuge, producing an oscillating Raman signal in the broad range of rotational frequencies. The observed periodic oscillations indicate coherent evolution of the excited rotational wave packets, with each Raman peak corresponding to the centrifuge-induced molecular alignment \cite{Korobenko2014a}.

One can see that the oscillation period remains relatively constant at lower rotation frequencies and then grows quickly as the frequency approaches critical values, specific to each particular molecule. Beyond this frequency, the rotational Raman signal disappears altogether, pointing to the fact that a molecule can no longer follow the centrifuge field, either because Coriolis force makes it fall behind the rotating trap (\nto{} and \ocs{}) or because of the rotation-induced dissociation (\cst{}). Supporting this conclusion is the observation of the Raman signal merging into a single line as the revival period diverges at the Coriolis wall (mostly pronounced in the \ocs{} spectrogram), which suggests that the rotational levels of \nto{} and \ocs{} become equidistant. In contrast to this behavior, the revival period of \cst{} remains finite even as the molecule falls out of the centrifuge due to dissociation.

The results of the quantitative analysis of $\Trev(\nu _{J})$ are shown in Fig.\ref{Fig-Analysis}. At lower frequencies, the experimental revival periods follow closely the dependencies derived from Eq.\ref{Eq-Trev} with $E(J)$ determined as the energy minimum of the corresponding Morse-cosine potential energy surface combined with the rotational term (Eq.\ref{Eq-Energy}). In this regime of negligible centrifugal distortion, the revival period is simply $\Trev=1/(4Bc)$ for asymmetric \nto{} and \ocs{} molecules, and $\Trev=1/(8Bc)$ for symmetric \cst{}, where $B$ is the corresponding rotational constant and $c$ is the speed of light in vacuum. The difference stems from the opposite contribution of the rotational wave packets with all-odd and all-even $J$ values, separately created by the centrifuge in an asymmetric molecule, to the Raman signal at $\Trev=1/(8Bc)$. On the contrary, the inversion symmetry and the nuclear spin statistics of \cst{} results in the absence of odd $J$'s in its rotational spectrum, and hence the appearance of fractional revivals with twice shorter period.

As the rotational frequency increases towards the upper frequency limits (dashed vertical lines), the experimentally measured revival periods begin to deviate from the calculated values. We attribute the disagreement to the zero-point stretching and bending, which have been neglected in our calculations. At extremely high levels of rotational excitation, the PES of a super-rotor becomes quite distorted and zero-point vibrations have to be taken into account for the correct calculation of the equilibrium bond lengths. Work is underway to verify whether this is the case, or whether Morse-cosine expansion of higher order is required in order to explain the observed molecular dynamics.

In summary, we presented a method of probing molecular potential energy surfaces with an optical centrifuge. Our approach is complementary to that based on vibrational spectroscopy, and offers much higher excitation energies. The method has been demonstrated with three linear molecules, \nto{}, \ocs{} and \cst{}. Ultra-fast rotation, induced by the centrifuge, caused the molecular bonds to stretch by as much as 32\%. In the case of \cst{}, our results suggest that the molecule dissociates after climbing to the extreme rotational level with a calculated angular momentum $\Jd=1186$. On the other hand, we confirmed experimentally that breaking molecular bonds with the centrifuge may not always be efficient. Incidentally, both \nto{} and \ocs{} cannot follow the accelerated rotation of the centrifuge field beyond the respective critical values $\Jc=460$ and $\Jc=650$, which fall short of the dissociation limits for each molecule. Similar analysis proved successful in describing the dynamics of centrifuged nonlinear sulfur dioxide \cite{Korobenko2016a}, and can be applied to any molecule amenable to centrifuge spinning, with an ultimate goal of selective bond breaking \cite{Hasbani2002}.

This research has been supported by the grants from CFI, BCKDF and NSERC and carried out under the auspices of the UBC Center for Research on Ultra-Cold Systems (CRUCS).


\begin{thebibliography}{16}%
\makeatletter
\providecommand \@ifxundefined [1]{%
 \@ifx{#1\undefined}
}%
\providecommand \@ifnum [1]{%
 \ifnum #1\expandafter \@firstoftwo
 \else \expandafter \@secondoftwo
 \fi
}%
\providecommand \@ifx [1]{%
 \ifx #1\expandafter \@firstoftwo
 \else \expandafter \@secondoftwo
 \fi
}%
\providecommand \natexlab [1]{#1}%
\providecommand \enquote  [1]{``#1''}%
\providecommand \bibnamefont  [1]{#1}%
\providecommand \bibfnamefont [1]{#1}%
\providecommand \citenamefont [1]{#1}%
\providecommand \href@noop [0]{\@secondoftwo}%
\providecommand \href [0]{\begingroup \@sanitize@url \@href}%
\providecommand \@href[1]{\@@startlink{#1}\@@href}%
\providecommand \@@href[1]{\endgroup#1\@@endlink}%
\providecommand \@sanitize@url [0]{\catcode `\\12\catcode `\$12\catcode
  `\&12\catcode `\#12\catcode `\^12\catcode `\_12\catcode `\%12\relax}%
\providecommand \@@startlink[1]{}%
\providecommand \@@endlink[0]{}%
\providecommand \url  [0]{\begingroup\@sanitize@url \@url }%
\providecommand \@url [1]{\endgroup\@href {#1}{\urlprefix }}%
\providecommand \urlprefix  [0]{URL }%
\providecommand \Eprint [0]{\href }%
\providecommand \doibase [0]{http://dx.doi.org/}%
\providecommand \selectlanguage [0]{\@gobble}%
\providecommand \bibinfo  [0]{\@secondoftwo}%
\providecommand \bibfield  [0]{\@secondoftwo}%
\providecommand \translation [1]{[#1]}%
\providecommand \BibitemOpen [0]{}%
\providecommand \bibitemStop [0]{}%
\providecommand \bibitemNoStop [0]{.\EOS\space}%
\providecommand \EOS [0]{\spacefactor3000\relax}%
\providecommand \BibitemShut  [1]{\csname bibitem#1\endcsname}%
\let\auto@bib@innerbib\@empty
\bibitem [{\citenamefont {Green}, \citenamefont {Moore},\ and\ \citenamefont
  {Polik}(1992)}]{Green1992}%
  \BibitemOpen
  \bibfield  {author} {\bibinfo {author} {\bibfnamefont {W.~H.}\ \bibnamefont
  {Green}}, \bibinfo {author} {\bibfnamefont {C.~B.}\ \bibnamefont {Moore}}, \
  and\ \bibinfo {author} {\bibfnamefont {W.~F.}\ \bibnamefont {Polik}},\
  }\href@noop {} {\bibfield  {journal} {\bibinfo  {journal} {Annual Review of
  Physical Chemistry}\ }\textbf {\bibinfo {volume} {43}},\ \bibinfo {pages}
  {591} (\bibinfo {year} {1992})}\BibitemShut {NoStop}%
\bibitem [{\citenamefont {Polanyi}\ and\ \citenamefont
  {Zewail}(1995)}]{Polanyi1995}%
  \BibitemOpen
  \bibfield  {author} {\bibinfo {author} {\bibfnamefont {J.~C.}\ \bibnamefont
  {Polanyi}}\ and\ \bibinfo {author} {\bibfnamefont {A.~H.}\ \bibnamefont
  {Zewail}},\ }\href@noop {} {\bibfield  {journal} {\bibinfo  {journal}
  {Accounts of Chemical Research}\ }\textbf {\bibinfo {volume} {28}},\ \bibinfo
  {pages} {119} (\bibinfo {year} {1995})}\BibitemShut {NoStop}%
\bibitem [{\citenamefont {Z\'{u}\~{n}iga}\ \emph {et~al.}(1999)\citenamefont
  {Z\'{u}\~{n}iga}, \citenamefont {Alacid}, \citenamefont {Bastida},
  \citenamefont {Carvajal},\ and\ \citenamefont {Requena}}]{Zuniga1999}%
  \BibitemOpen
  \bibfield  {author} {\bibinfo {author} {\bibfnamefont {J.}~\bibnamefont
  {Z\'{u}\~{n}iga}, \bibfnamefont {J.}}, \bibinfo {author} {\bibfnamefont
  {M.}~\bibnamefont {Alacid}}, \bibinfo {author} {\bibfnamefont
  {A.}~\bibnamefont {Bastida}}, \bibinfo {author} {\bibfnamefont {F.~J.}\
  \bibnamefont {Carvajal}}, \ and\ \bibinfo {author} {\bibfnamefont
  {A.}~\bibnamefont {Requena}},\ }\href@noop {} {\bibfield  {journal} {\bibinfo
   {journal} {The Journal of Chemical Physics}\ }\textbf {\bibinfo {volume}
  {110}},\ \bibinfo {pages} {6339} (\bibinfo {year} {1999})}\BibitemShut
  {NoStop}%
\bibitem [{\citenamefont {Z\'{u}\~{n}iga}\ \emph {et~al.}(2000)\citenamefont
  {Z\'{u}\~{n}iga}, \citenamefont {Bastida}, \citenamefont {Alacid},\ and\
  \citenamefont {Requena}}]{Zuniga2000}%
  \BibitemOpen
  \bibfield  {author} {\bibinfo {author} {\bibfnamefont {J.}~\bibnamefont
  {Z\'{u}\~{n}iga}}, \bibinfo {author} {\bibfnamefont {A.}~\bibnamefont {Bastida}},
  \bibinfo {author} {\bibfnamefont {M.}~\bibnamefont {Alacid}}, \ and\ \bibinfo
  {author} {\bibfnamefont {A.}~\bibnamefont {Requena}},\ }\href@noop {}
  {\bibfield  {journal} {\bibinfo  {journal} {The Journal of Chemical Physics}\
  }\textbf {\bibinfo {volume} {113}},\ \bibinfo {pages} {5695} (\bibinfo {year}
  {2000})}\BibitemShut {NoStop}%
\bibitem [{\citenamefont {Z\'{u}\~{n}iga}\ \emph {et~al.}(2002)\citenamefont
  {Z\'{u}\~{n}iga}, \citenamefont {Bastida}, \citenamefont {Requena},\ and\
  \citenamefont {Sibert}}]{Zuniga2002}%
  \BibitemOpen
  \bibfield  {author} {\bibinfo {author} {\bibfnamefont {J.}~\bibnamefont
  {Z\'{u}\~{n}iga}}, \bibinfo {author} {\bibfnamefont {A.}~\bibnamefont {Bastida}},
  \bibinfo {author} {\bibfnamefont {A.}~\bibnamefont {Requena}}, \ and\
  \bibinfo {author} {\bibfnamefont {E.~L.}\ \bibnamefont {Sibert}},\
  }\href@noop {} {\bibfield  {journal} {\bibinfo  {journal} {The Journal of
  Chemical Physics}\ }\textbf {\bibinfo {volume} {116}},\ \bibinfo {pages}
  {7495} (\bibinfo {year} {2002})}\BibitemShut {NoStop}%
\bibitem [{\citenamefont {Jensen}(1988)}]{Jensen1988}%
  \BibitemOpen
  \bibfield  {author} {\bibinfo {author} {\bibfnamefont {P.}~\bibnamefont
  {Jensen}},\ }\href@noop {} {\bibfield  {journal} {\bibinfo  {journal}
  {Journal of Molecular Spectroscopy}\ }\textbf {\bibinfo {volume} {128}},\
  \bibinfo {pages} {478} (\bibinfo {year} {1988})}\BibitemShut {NoStop}%
\bibitem [{\citenamefont {Karczmarek}\ \emph {et~al.}(1999)\citenamefont
  {Karczmarek}, \citenamefont {Wright}, \citenamefont {Corkum},\ and\
  \citenamefont {Ivanov}}]{Karczmarek1999}%
  \BibitemOpen
  \bibfield  {author} {\bibinfo {author} {\bibfnamefont {J.}~\bibnamefont
  {Karczmarek}}, \bibinfo {author} {\bibfnamefont {J.}~\bibnamefont {Wright}},
  \bibinfo {author} {\bibfnamefont {P.}~\bibnamefont {Corkum}}, \ and\ \bibinfo
  {author} {\bibfnamefont {M.~Yu.}~\bibnamefont {Ivanov}},\ }\href@noop {}
  {\bibfield  {journal} {\bibinfo  {journal} {Physical Review Letters}\
  }\textbf {\bibinfo {volume} {82}},\ \bibinfo {pages} {3420} (\bibinfo {year}
  {1999})}\BibitemShut {NoStop}%
\bibitem [{\citenamefont {Villeneuve}\ \emph {et~al.}(2000)\citenamefont
  {Villeneuve}, \citenamefont {Aseyev}, \citenamefont {Dietrich}, \citenamefont
  {Spanner}, \citenamefont {Ivanov},\ and\ \citenamefont
  {Corkum}}]{Villeneuve2000}%
  \BibitemOpen
  \bibfield  {author} {\bibinfo {author} {\bibfnamefont {D.~M.}\ \bibnamefont
  {Villeneuve}}, \bibinfo {author} {\bibfnamefont {S.~A.}\ \bibnamefont
  {Aseyev}}, \bibinfo {author} {\bibfnamefont {P.}~\bibnamefont {Dietrich}},
  \bibinfo {author} {\bibfnamefont {M.}~\bibnamefont {Spanner}}, \bibinfo
  {author} {\bibfnamefont {M.~Yu.}\ \bibnamefont {Ivanov}}, \ and\ \bibinfo
  {author} {\bibfnamefont {P.~B.}\ \bibnamefont {Corkum}},\ }\href@noop {}
  {\bibfield  {journal} {\bibinfo  {journal} {Physical Review Letters}\
  }\textbf {\bibinfo {volume} {85}},\ \bibinfo {pages} {542} (\bibinfo {year}
  {2000})}\BibitemShut {NoStop}%
\bibitem [{\citenamefont {Campargue}\ \emph {et~al.}(1995)\citenamefont
  {Campargue}, \citenamefont {Permogorov}, \citenamefont {Bach}, \citenamefont
  {Temsamani}, \citenamefont {Vander~Auwera}, \citenamefont {Herman},\ and\
  \citenamefont {Fujii}}]{Campargue1995}%
  \BibitemOpen
  \bibfield  {author} {\bibinfo {author} {\bibfnamefont {A.}~\bibnamefont
  {Campargue}}, \bibinfo {author} {\bibfnamefont {D.}~\bibnamefont
  {Permogorov}}, \bibinfo {author} {\bibfnamefont {M.}~\bibnamefont {Bach}},
  \bibinfo {author} {\bibfnamefont {A.}~\bibnamefont {Temsamani}}, \bibinfo
  {author} {\bibfnamefont {J.}~\bibnamefont {Vander~Auwera}}, \bibinfo {author}
  {\bibfnamefont {M.}~\bibnamefont {Herman}}, \ and\ \bibinfo {author}
  {\bibfnamefont {M.}~\bibnamefont {Fujii}},\ }\href@noop {} {\bibfield
  {journal} {\bibinfo  {journal} {The Journal of Chemical Physics}\ }\textbf
  {\bibinfo {volume} {103}},\ \bibinfo {pages} {5931} (\bibinfo {year}
  {1995})}\BibitemShut {NoStop}%
\bibitem [{\citenamefont {Spanner}\ and\ \citenamefont
  {Ivanov}(2001)}]{Spanner2001}%
  \BibitemOpen
  \bibfield  {author} {\bibinfo {author} {\bibfnamefont {M.}~\bibnamefont
  {Spanner}}\ and\ \bibinfo {author} {\bibfnamefont {M.~Yu.}\ \bibnamefont
  {Ivanov}},\ }\href@noop {} {\bibfield  {journal} {\bibinfo  {journal} {The
  Journal of Chemical Physics}\ }\textbf {\bibinfo {volume} {114}},\ \bibinfo
  {pages} {3456} (\bibinfo {year} {2001})}\BibitemShut {NoStop}%
\bibitem [{\citenamefont {Hasbani}\ \emph {et~al.}(2002)\citenamefont
  {Hasbani}, \citenamefont {Ostojic}, \citenamefont {Bunker},\ and\
  \citenamefont {Ivanov}}]{Hasbani2002}%
  \BibitemOpen
  \bibfield  {author} {\bibinfo {author} {\bibfnamefont {R.}~\bibnamefont
  {Hasbani}}, \bibinfo {author} {\bibfnamefont {B.}~\bibnamefont {Ostojic}},
  \bibinfo {author} {\bibfnamefont {P.~R.}\ \bibnamefont {Bunker}}, \ and\
  \bibinfo {author} {\bibfnamefont {M.~Yu.}\ \bibnamefont {Ivanov}},\
  }\href@noop {} {\bibfield  {journal} {\bibinfo  {journal} {The Journal of
  Chemical Physics}\ }\textbf {\bibinfo {volume} {116}},\ \bibinfo {pages}
  {10636} (\bibinfo {year} {2002})}\BibitemShut {NoStop}%
\bibitem [{\citenamefont {Korobenko}, \citenamefont {Milner},\ and\
  \citenamefont {Milner}(2014)}]{Korobenko2014a}%
  \BibitemOpen
  \bibfield  {author} {\bibinfo {author} {\bibfnamefont {A.}~\bibnamefont
  {Korobenko}}, \bibinfo {author} {\bibfnamefont {A.~A.}\ \bibnamefont
  {Milner}}, \ and\ \bibinfo {author} {\bibfnamefont {V.}~\bibnamefont
  {Milner}},\ }\href@noop {} {\bibfield  {journal} {\bibinfo  {journal}
  {Physical Review Letters}\ }\textbf {\bibinfo {volume} {112}},\ \bibinfo
  {pages} {113004} (\bibinfo {year} {2014})}\BibitemShut {NoStop}%
\bibitem [{\citenamefont {Leichtle}, \citenamefont {Averbukh},\ and\
  \citenamefont {Schleich}(1996)}]{Leichtle1996}%
  \BibitemOpen
  \bibfield  {author} {\bibinfo {author} {\bibfnamefont {C.}~\bibnamefont
  {Leichtle}}, \bibinfo {author} {\bibfnamefont {I.~Sh.}\ \bibnamefont
  {Averbukh}}, \ and\ \bibinfo {author} {\bibfnamefont {W.~P.}\ \bibnamefont
  {Schleich}},\ }\href@noop {} {\bibfield  {journal} {\bibinfo  {journal}
  {Physical Review Letters}\ }\textbf {\bibinfo {volume} {77}},\ \bibinfo
  {pages} {3999} (\bibinfo {year} {1996})}\BibitemShut {NoStop}%
\bibitem [{\citenamefont {Seideman}(1999)}]{Seideman1999}%
  \BibitemOpen
  \bibfield  {author} {\bibinfo {author} {\bibfnamefont {T.}~\bibnamefont
  {Seideman}},\ }\href@noop {} {\bibfield  {journal} {\bibinfo  {journal}
  {Physical Review Letters}\ }\textbf {\bibinfo {volume} {83}},\ \bibinfo
  {pages} {4971} (\bibinfo {year} {1999})}\BibitemShut {NoStop}%
\bibitem [{\citenamefont {Averbukh}\ and\ \citenamefont
  {Perelman}(1989)}]{Averbukh1989}%
  \BibitemOpen
  \bibfield  {author} {\bibinfo {author} {\bibfnamefont {I.~Sh.}\ \bibnamefont
  {Averbukh}}\ and\ \bibinfo {author} {\bibfnamefont {N.~F.}\ \bibnamefont
  {Perelman}},\ }\href@noop {} {\bibfield  {journal} {\bibinfo  {journal}
  {Physics Letters A}\ }\textbf {\bibinfo {volume} {139}},\ \bibinfo {pages}
  {449} (\bibinfo {year} {1989})}\BibitemShut {NoStop}%
\bibitem [{\citenamefont {Korobenko}\ and\ \citenamefont
  {Milner}(2016)}]{Korobenko2016a}%
  \BibitemOpen
  \bibfield  {author} {\bibinfo {author} {\bibfnamefont {A.}~\bibnamefont
  {Korobenko}}\ and\ \bibinfo {author} {\bibfnamefont {V.}~\bibnamefont
  {Milner}},\ }\href@noop {} {\bibfield  {journal} {\bibinfo  {journal}
  {Physical Review Letters}\ }\textbf {\bibinfo {volume} {116}},\ \bibinfo
  {pages} {183001} (\bibinfo {year} {2016})}\BibitemShut {NoStop}%
\end{thebibliography}

%

\end{document}